\documentclass{cimento}

\usepackage{graphicx}  

\usepackage{amsmath, amssymb}     
\usepackage{booktabs}
\usepackage{fancyhdr}
\usepackage{float}

\renewcommand{\shorttitle}{HE Neutrino Flux from SN2024ggi}

\title{High-energy neutrino flux from SN2024ggi: constraints from semi-analytic modeling of its post-explosive emission}
\author{
M.~Buccheri\from{ins:b}\from{ins:a}\from{ins:y}\ETC,
S.~P.~Cosentino\from{ins:x}\from{ins:y},
M.~L.~Pumo\from{ins:x}\from{ins:y}\from{ins:l}
}

\instlist{
  \inst{ins:b} INAF — Osservatorio Astronomico d'Abruzzo — Teramo, Italy
  \inst{ins:x} INAF — Osservatorio Astrofisico di Catania — Catania, Italy
  \inst{ins:a} Università degli Studi di Roma “Tor Vergata”, Dipartimento di Fisica — Roma, Italy
  \inst{ins:y} Università di Catania, Dip. di Fisica e Astronomia “Ettore Majorana” — Catania, Italy
  \inst{ins:l} Laboratori Nazionali del Sud, INFN — Catania, Italy
}

\begin{document}

\maketitle

\begin{abstract}
Hydrogen-rich supernovae can efficiently accelerate particles when the expanding ejecta interact with the surrounding circumstellar medium (CSM), producing high-energy (TeV--PeV) neutrinos. In this work we investigate the nearby SN~2024ggi, whose proximity and clear signatures of ejecta--CSM interaction make it a promising candidate for studying high-energy neutrino (HE-$\nu$) emission. We apply a new semi-analytical model that consistently links the electromagnetic and neutrino emission components, allowing us to constrain the main explosion parameters, including the kinetic energy, ejecta mass, progenitor radius, and nickel yield. The predicted HE-$\nu$ fluence at Earth peaks at TeV energies and remains below the sensitivity of current detectors. However, the modeling establishes a robust framework for interpreting future signals from nearby interacting supernovae and fines tune observational strategies for next-generation multi-messenger facilities such as IceCube-Gen2 and KM3NeT/ARCA.
 
\end{abstract}

\section{Introduction}
Supernovae (SNe) mark the catastrophic end of massive stars, typically with Zero-Age Main Sequence (ZAMS) masses larger than $\sim 8$--$10\,M_\odot$ \cite{ref:Woosley2002}. These explosions release enormous amounts of energy, on the order of $\sim 10^{51}$~erg, and enrich the surrounding environment with newly synthesized material. Their peak luminosities can reach $\sim 10^{41}$--$10^{42}$~erg~s$^{-1}$, comparable to that of their host galaxies ($\sim 10^{8}$--$10^{9}\,L_\odot$; see Fig.~1).
Among them, hydrogen-rich SNe are particularly significant, as the interaction between the SN ejecta and the CSM can serve as an efficient site for particle acceleration up to very high energies. In this framework, inelastic proton–proton collisions naturally lead to the production of HE-$\nu$, which are invaluable messengers in astrophysics: owing to their weakly interacting nature, they can traverse cosmic distances essentially unimpeded, carrying direct information from their regions of origin.
SN 2024ggi, located at a relatively small distance of about 6.7 Mpc \cite{ref:Chen2025}, exhibits clear signatures of ejecta–CSM interaction from the early stages. This makes it an ideal case study for jointly investigating electromagnetic and HE-$\nu$ emission, with the aim of constraining both the properties of the progenitor system and the structure of its circumstellar environment. Understanding these processes also provides important insights into particle-acceleration mechanisms in core-collapse (CC) events, thereby contributing to the broader context of high-energy (HE) astrophysics.

\begin{figure}[h!]
  \centering
  \includegraphics[scale=0.41]{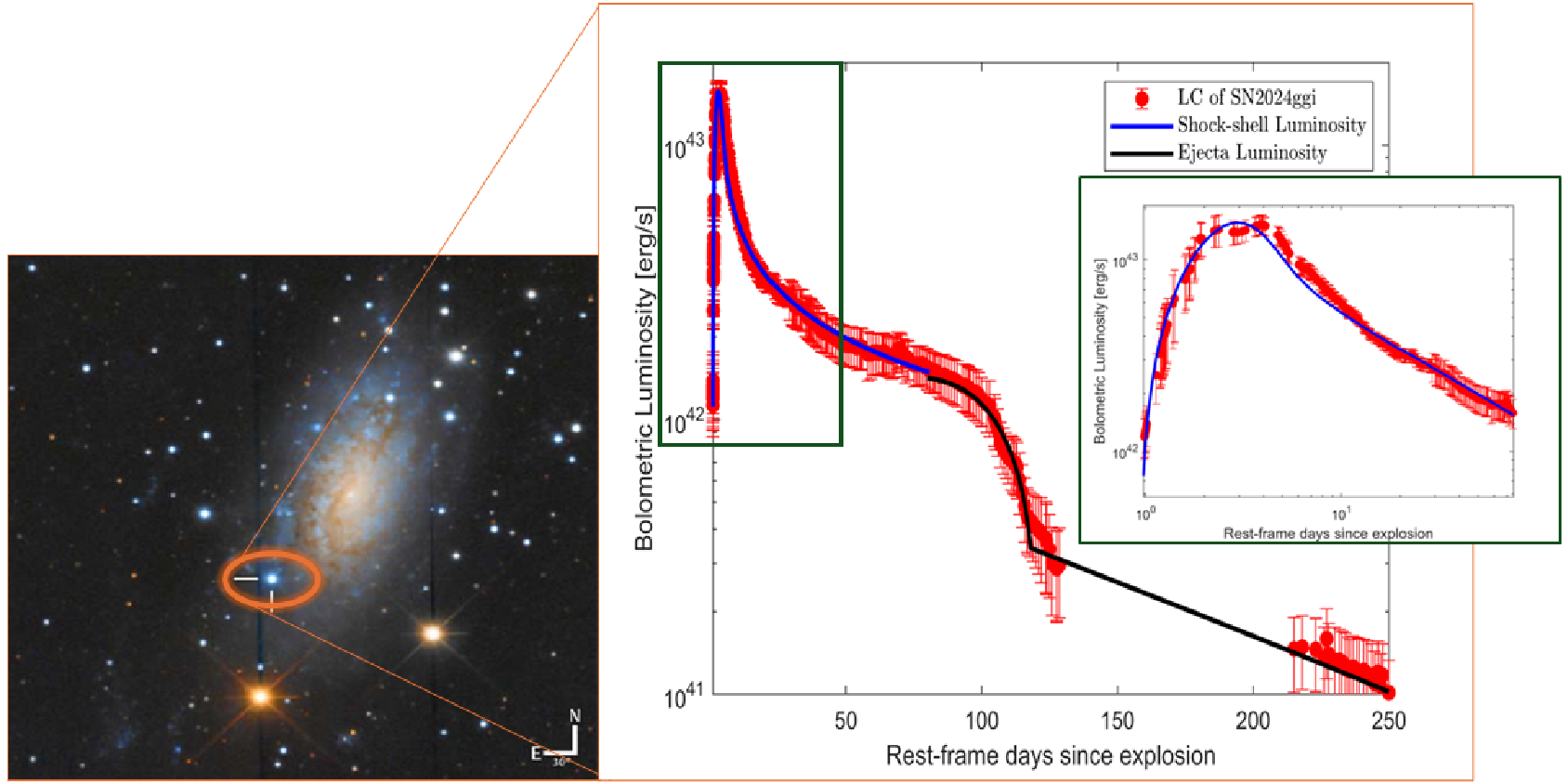}
  \caption{SN 2024ggi in NGC 3621 from \cite{ref:Chen2025}. The plots show the bolometric light curve analyzed in this work, with model curves for shock emission and the ejecta after emerging from the CSM.}

  \label{Fig:SN}
\end{figure}

\section{HE $\nu$ Emission from Core-Collapse Supernovae}

According to standard stellar evolution theory, massive stars complete all nuclear burning stages up to the formation of an iron core, which becomes gravitationally unstable and collapses. During the collapse, an intense flux of neutrinos is emitted from the core, depositing only $\sim 1\%$ of its energy in the stellar envelope \cite{ref:Janka}. These MeV neutrinos provide the only direct observational probe of the CC “explosive” phase.
After the explosion, the fast-expanding ejecta interact with the surrounding CSM, generating a delayed HE-$\nu$ emission. The interaction drives two shock fronts: a forward shock propagating through the CSM and a reverse shock moving inwards in mass. While both shocks contribute to particle acceleration, efficient energy gain for protons occurs primarily in the forward shock propagating in optically thin, low-density CSM regions. Inelastic collisions between these accelerated protons and the swept-up CSM nuclei produce $\eta$ and $\pi$ mesons, which subsequently decay into HE-$\nu$ and gamma rays. The production of HE-$\nu$ can be quantitatively described in terms of the proton population within the forward shock \cite{ref:Cosentino}.The differential HE-$\nu$ production rate of flavour $i$ at the source, at time $t$ and HE-$\nu$ energy $E_\nu$, is then given by:
\begin{equation}
Q_{\nu_i + \bar{\nu}_i} (E_\nu, t) = \frac{c \, n_{\rm sh}[R_{\rm sh}(t)]}{0.938~{\rm GeV}} \; \bar{Q}_{\nu_i + \bar{\nu}_i}(E_\nu, t),
\end{equation}
where $n_{\rm sh}$ is the proton density in the shocked shell, $R_{\rm sh}(t)$ the shock radius, and $\bar{Q}_{\nu_i + \bar{\nu}_i}$ accounts for pp-collision physics and secondary particle decay \cite{ref:Cosentino}.
Accounting for distance $D$, redshift $z$, and flavour oscillations (via $P_{\nu_\beta \to \nu_\alpha}$ \cite{ref:Cosentino}), the flux observed at Earth is:
\begin{equation}
F_{\nu_\alpha + \bar{\nu}_\alpha}(E_\nu, t) = \sum_\beta \frac{P_{\nu_\beta \to \nu_\alpha}}{4 \pi D^2} \; \times Q_{\nu_\beta + \bar{\nu}_\beta}[E_\nu (1+z), t/(1+z)].
\end{equation}
Finally, integrating over the duration of the shock-CSM interaction, i.e from the breakout time $t_{\rm bo}$ to the outer boundary of the CSM $t_{\rm f}$, gives the total time-averaged flux (fluence):
\begin{equation}
\Phi_{\nu_\mu}(E_\nu) = \frac{1}{t_{\rm f} - t_{\rm bo}} \int_{t_{\rm bo}}^{t_{\rm f}} F_{\nu_\mu + \bar{\nu}_\mu}(E_\nu, t) \, dt.
\end{equation}


\section{Case of SN~2024ggi}

SN~2024ggi was discovered by ATLAS on 2024 April 11 in the nearby spiral galaxy NGC~3621 (RA = $11^{\text{h}}18^{\text{m}}22.087^{\text{s}}$, Dec = $-32^{\circ}50'15.27'')$, located in the southern outskirts of the galaxy \cite{ref:Chen2025, ref:Tonry2018}.Classified as a Type IIP SN, it had a plateau bolometric luminosity $\sim 10^{42}$erg $s^{-1}$ extending to $\sim 120$ days after explosion. Early flash spectroscopy revealed high-ionization emission lines persisting for $\sim$3.8 days, signaling interaction with a CSM. The bolometric light curve (LC; see Fig.1), constructed from UV-to-near-IR photometry with extinction and distance corrections, is interpreted assuming full gamma-ray trapping from the $^{56}\mathrm{Ni} \rightarrow {}^{56}\mathrm{Co}$ decay chain. The collected data, from $\sim 2$ to $\sim$250 days after the explosion, show a bolometric lumonisity peak of $\sim 10^{43}$ erg s$^{-1}$ due to shock breakout and interaction with a low-density, extended CSM, followed by a hydrogen-recombination plateau and a late-time decline tracing $^{56}$Ni decay, allowing a robust estimate of the $^{56}$Ni mass. Through modeling of the LC, we obtained the key explosion parameters reported in Tab. \ref{tab:sn2024ggi_params}, including the kinetic energy ($E_k$), mass ($M_{ej}$) and initial radius ($R_\star$) of the ejecta, the ejected $^{56}$Ni mass ($M_{Ni}$) and the CSM density properties, such as mass ($M_{CSM}$), radius ($R_{CSM}$) and density slope ($s$).

\begin{table}[H]
    \centering
    \caption{Preliminary parameters of SN 2024ggi derived from the LC modeling.}
    \label{tab:sn2024ggi_params}
    \begin{tabular}{llcl}
        \toprule
        \multicolumn{4}{c}{Explosion and CSM parameters of SN 2024ggi} \\
        \midrule
        $E_k$      & $1.1 \pm 0.2$ foe           & $R_{\mathrm{CSM}}$ & $(4.5 \pm 0.5) \times 10^{15}$ cm \\
        $M_{\mathrm{ej}}$ & $13 \pm 0.5~M_\odot$ & $M_{\mathrm{CSM}}$ & $5^{+2}_{-1} \times 10^{-2}~M_\odot$ \\
        $R_\star$  & $(2 \pm 0.6) \times 10^{13}$ cm & $s$ & $2.35 \pm 0.1$ \\
        $M_{\mathrm{Ni}}$ & $6.5^{+1}_{-0.15} \times 10^{-2}~M_\odot$ \\

        \bottomrule
    \end{tabular}
\end{table}

Using the best-fit ejecta and CSM parameters, the HE-$\nu$ emission from SN~2024ggi was computed for the shock–CSM interaction phase \cite{ref:Cosentino}. The differential muon–neutrino flux ($E_\nu^2\,F_{\nu_\mu+\bar{\nu}_\mu}$) peaks at $E_\nu \sim 1$ TeV with a maximum value of $\sim 4 \times 10^{-9}$ GeV cm$^{-2}$ s$^{-1}$ during the first $\sim$2 days, when the shock traverses the densest CSM layers. As the ejecta expand and the surrounding density decreases, the flux drops by roughly one order of magnitude over $\sim$45 days. The resulting time-integrated fluence is $ \sim 2 \times 10^{-4}$ erg cm$^{-2}$, dominated by TeV energies with rapid suppression above PeV scales. This corresponds to an expected number of $\nu_\mu$ events at IceCube of $N_{\nu_\mu} \sim 10^{-4}$ for neutrino energy above $0.1$ TeV.  
Despite its relatively short distance, the expected HE-$\nu$ signal from SN~2024ggi at IceCube remains well below the background fluctuations. Its southern declination ($\delta_\star \simeq -32.8^\circ$) reduces the expected $\nu_\mu$ rate due to IceCube’s low effective area in the southern sky; at a more favorable declination, such as SN~2023ixf ($\delta_\star \simeq 54.3^\circ$; \cite{ref:Cosentino}), the expected detection rate would be more than two orders of magnitude higher. These results highlight the importance of complementary Northern Hemisphere detectors, such as KM3NeT/ARCA, for observing Southern-sky CC-SNe.

\section{Considerations and Further Comments}

Interacting SNe like SN 2024ggi provide a unique laboratory for HE astrophysics. Modeling the bolometric LC of SN 2004ggi allowed us to constrain the main properties of its progenitor at the explosion as well as the CSM structure, which were then used for HE-$\nu$ emission calculations. Our results indicate that the HE-$\nu$ output is concentrated within the first two days after the explosion and is dominated by TeV energies. Despite its proximity, the expected signal at IceCube remains well below the detector sensitivity due to its southern declination, emphasizing the importance of Northern Hemisphere detectors such as KM3NeT/ARCA for Southern-sky events and the use of increasingly sensitive detectors  \cite{ref:IceCube2021,ref:KM3NeT2017}. From an astrophysical perspective, it is useful to refine the models used to describe these events, in order to achieve more accurate predictions of HE-$\nu$ fluxes and better define the temporal window during which HE-$\nu$ emission can be detected. Future advancements should incorporate a more realistic treatment of the ejecta and CSM density profiles, accounting for deviations from simple power-law prescriptions, radial inhomogeneities, and potential asymmetries. Additionally, extending observational coverage beyond 250 days would provide tighter constraints on the late-time evolution of the bolometric LC and the onset of radioactive gamma-ray emission resulting from Ni decay leakage. A more detailed approach to gamma-ray transport, including partial trapping at late times, is therefore crucial for accurately interpreting the radioactive tail, as non-local trapping and leakage become significant when ejecta density decreases and the gamma-ray optical depth approaches unity \cite{ref:Zampieri2017}. All these improvements would also facilitate statistically meaningful constraints across a broader sample of interacting CC-SNe. Indeed, events like SN 2024ggi and similar transients are prime targets for multimessenger astrophysics, connecting detailed optical observations with HE-$\nu$ studies and probing the most extreme natural particle accelerators in the Universe. \\

\acknowledgments
We acknowledge support from the Piano di Ricerca di Ateneo UNICT – Linea 2 PIA.CE.RI. 2024–2026 of Catania University (project AstroCosmo, P.I. A. Lanzafame).

\end{document}